\documentclass[10pt,letter]{elsart}

\usepackage{epsf}
\usepackage{epsfig}
\usepackage{float}
\usepackage{subfigure}
\usepackage{verbatim}
\usepackage{amsmath}
\usepackage{caption}

\begin{document}

\newcommand\at[2]{\left.#1\right|_{#2}}
\newcommand{\He}{\nuc{3}{He}}
\newcommand{\Hy}{\nuc{1}{H}}
\newcommand{\D}{\nuc{2}{H}}
\newcommand{\Trit}{\nuc{3}{H}}

\vspace*{-0.5cm}
\begin{frontmatter}
\title{\bf\normalsize{E12-14-009: Ratio of the electric form factor \\in the mirror nuclei \He\ and \Trit\ } }
\vspace*{-1.0cm}
\footnotesize\author{L.~S.~Myers\thanksref{contact}}
\author{ and D.~W.~Higinbotham}
\address{Thomas Jefferson National Accelerator Facility,
Newport News, VA 23606, USA}
\footnotesize\author{J.~R.~Arrington}
\address{Argonne National Laboratory, Argonne, IL 60439, USA}

\thanks[contact]{contact person: email \texttt{lmyers$@$jlab.org} }

\begin{abstract}
\label{abstract}
\footnotesize
E12-14-009: We propose to extract the ratio of the electric form factor ($G_E$) of \He\ and \Trit\ from the measured ratio of the elastic-scattering cross sections at $E_{\rm beam}$~=~1.1~GeV. Measurements at low $Q^2$ ( $\le$ 0.1 GeV$^2$) will allow accurate extraction of $G_E$ with minimal contributions from the magnetic form factor ($G_M$) and Coulomb corrections. From this data we will extract the difference between the charge radii for \He\ and \Trit. This short experiment, 1.5 days, will utilize the left Hall A high resolution spectrometer and the one-time availability of a 1~kCi \Trit\ target at Jefferson Lab which has been approved for the E12-10-103, E12-11-112 and E12-14-011 experiments.

\end{abstract}

\end{frontmatter}
\section{Introduction}
\label{intro}

The electric ($G_E$) and magnetic ($G_M$) form factors of nucleons and nuclei can be accessed via electron scattering at low $Q^2$ from the object in question. In the case of a nucleon, the scattering cross section is given by the Rosenbluth formula
\begin{equation}
  \label{eq:eq2}
  \frac{d\sigma}{d\Omega} = \sigma_{\rm Mott} \left[G_{\rm E}^2 + \frac{\tau}{\varepsilon}G_{\rm M}^2\right] (1+\tau)^{-1},
\end{equation} 
where $\sigma_{\rm Mott}$ is the cross section for scattering from a point nucleon, and $\tau$=$Q^2$/$4M^2$ and $\varepsilon^{-1}$=\{$1+2(1+\tau){\rm tan}^2 \theta/2$\} are kinematic factors. At low $Q^2$, $\tau \ll$ 1 and $\varepsilon \approx$ 1, so that the cross section is dominated by $G_E$ and is mostly insensitive to $G_M$. 

The electric (magnetic) form factor is related to the electric (magnetic) charge distribution, $\rho_{\rm E (M)}$, via
\begin{equation}
  \label{eq:eq3}
  \begin{split}
  G_{\rm E(M)} & = \int \rho_{\rm E(M)} ({\rm \vec{r}}) e^{i {\rm \vec{q} \cdot \vec{r} } } d^{\rm 3}{\rm \vec{r}} \\
              & \approx 1 - \frac{1}{6} \vec{q}^2 \langle r^2_{\rm E(M)} \rangle +  \cdots
  \end{split}
\end{equation} 
where the charge density is assumed to be spherically symmetric, $\langle r^2 \rangle$ is the charge radius and higher-order terms have been omitted. By measuring the form factor at low momentum transfer, $Q^2(=-q^2)$, one can extract the charge radius from the slope of the form factor,
\begin{equation}
  \label{eq:eq1}
  \langle r^2_{\rm E} \rangle = -6 \at{\frac{dG_{\rm E}}{dQ^2}}{Q = 0}.
\end{equation} 

The first \Trit/\He($e$,$e^\prime$) measurements were performed nearly fifty years ago at SLAC \cite{collard65}. Since then, many studies of \He\ \cite{mccarthy77,arnold78,szalata77,ottermann85,dunn83,beck87,amroun94} and  \Trit\ \cite{juster85,beck84,beck87,amroun94} have been conducted at $Q^2~\le$~1~GeV$^2$. The best, and most recent of the \Trit\ experiments are from Bates \cite{beck87} and SACLAY \cite{amroun94}, and in both experiments measurements of \He\ were also taken. The extraction of both radii by the Bates group are $\sim$0.1~fm smaller than the SACLAY group (see Table~\ref{table4}), although the Bates fits typically have large values of $\chi^2$. The SACLAY results are the ones most often cited and so we will use those for the remainder of the proposal.

\begin{table}
\begin{center}
\caption{Charge radii (in fm) for \Trit\ and \He.}
\begin{tabular}{lcc}
\hline \hline
Ref. & \Trit\ & \He\ \\
SACLAY\cite{amroun94} & 1.76 $\pm$ 0.09 & 1.96 $\pm$ 0.03 \\
Bates only\cite{beck84}  &  1.68 $\pm$ 0.03 &  1.87 $\pm$ 0.03 \\
GFMC \cite{liphd} & 1.77 $\pm$ 0.01 & 1.97 $\pm$ 0.01 \\
$\chi$EFT \cite{piarulli13} & 1.756 $\pm$ 0.006 & 1.962 $\pm$ 0.004 \\
\hline \hline
\end{tabular}
  \label{table4}
\end{center}
\end{table}

Measurements of the charge radii for \Trit\ and \He, assuming isospin symmetry, allow for the separation of the proton and neutron radius distributions. The separation is sensitive to differences between the p-p, n-p, and n-n interactions, as well as isospin-dependence in the three-body force.  These results, in terms of \Trit\ and \He\ (or proton and neutron) radii, can be compared against precise ab initio calculations using phenomenological or chiral N-N potentials.

The last measurement of the \Trit\ charge radius was made twenty years ago. Recent developments in Chiral Effective Field Theory ($\chi$EFT) \cite{piarulli13} and  Green’s function Monte Carlo methods (GFMC) \cite{liphd}, though, have demonstrated that the form factors and charge radii can be calculated extremely accurately. These calculations require testing by even-more-precise experimental results.

A new measurement of the \He\ charge radius has been made utilizing the isotope shift of spectral lines in Helium \cite{morton06} that is more than an order of magnitude more precise than the value obtained from the world elastic scattering data \cite{amroun94} (Table~\ref{table5}). This measurement (and others like it) rely on a precise extraction of the radius from another technique ({\it e.g.} electron scattering) in order for proper normalization. A measurement of the relative \Trit, \He\ radii will allow for a more precise connection between the hydrogen isotope chain and the helium isotope chain.\footnote{A measurement of the \He\ radius using the Lamb shift of muonic \He\ is planned by the CREMA collaboration. This will provide an independent connection between the radii of hydrogen and \He, which can may provide an additional check on the consistency between the muonic hydrogen extractions and other measurements.} In addition, a measurement of the electric form factors of \Trit\ and \He, combined with the known \He\ radius, would produce a more precise measure of the \Trit\ charge radius which could then be compared to theoretical calculations. 

\begin{table}
\begin{center}
\caption{Theoretical calculation and experimental results of rms nuclear charge radii (in fm). Adapted from \cite{liphd} -- for individual references see \cite{amroun94,pieper1,sick03,sick96,sick82,alkhazov97,melnikov00,huber98,shiner95,borie78}. The majority of the atomic measurements are isotope shifts which yield the difference in radius between nuclei. These require a single absolute measurement to determine the radius of all of the hydrogen (helium) isotopes.}
\begin{tabular}{lcccc}
\hline \hline
 & Theory &  ($e$,$e^\prime$) & ($p$,$p^\prime$) & Atomic Measurements \\
\hline
\nuc{1}{H} & --- & 0.895(18) & --- & 0.883(14) \\
\nuc{2}{H} & 2.14(1) & 2.128(11) & --- & 2.145(6) \\
\nuc{3}{H} & 1.77(1) & 1.755(86) & --- & --- \\
\nuc{3}{He} & 1.97(1) & 1.959(30) & --- & 1.9506(14) \\
\nuc{4}{He} & 1.68(1) & 1.676(8) & 1.71(3) & 1.673(1) \\
\nuc{6}{He} & 2.06(1) & --- & 2.03(11) & 2.054(14) \\
\hline \hline
\end{tabular}
  \label{table5}
\end{center}
\end{table}

\section{Proposed measurement}

\subsection{Kinematics and Equipment}
We propose to measure the ratio of the \Trit:\He\ elastic cross section at a beam energy of $\sim$1.1~GeV. Scattering data at $Q^2$~=~0.05--0.09~GeV$^2$ will be obtained by using the left High Resolution Spectrometer (HRS) in Hall~A placed at scattering angles of 12.5$^\circ$ and 15.0$^\circ$. The HRS angle is the only variable in the experiment that will be changed. The HRS momentum acceptance ($\pm$4\%) is large enough to accommodate elastically-scattered electrons at both angles. This reduces the systematic uncertainty in the ratio as well as making the measurement more straightforward.

\begin{figure}
\setlength{\epsfxsize}{0.70\textwidth}
\leavevmode
\begin{center}
\epsffile{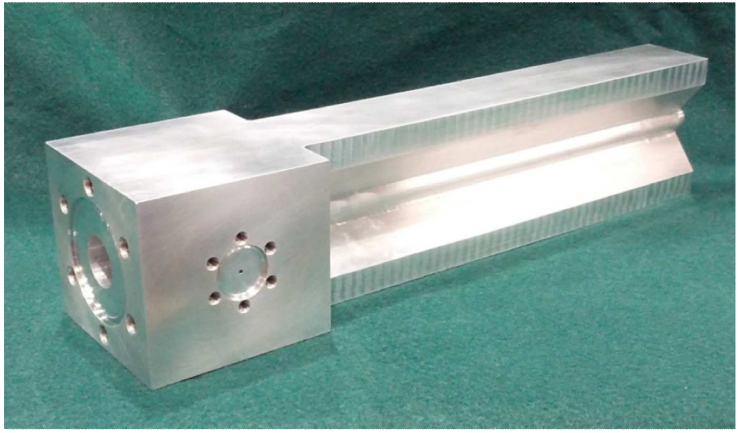}
\end{center}
\caption{Picture of the target cell for the E12-10-103, E12-11-112, and E12-14-011 experiments.}
\label{fig0}
\end{figure}

The \nuc{3}{He} and \nuc{3}{H} targets will be the same as the ones used in the approved E12-10-103, E12-11-112, and E12-14-011 experiments (Fig.~\ref{fig0}) \cite{exp1,exp2,exp3}, thus eliminating the biggest hurdle to completing the measurement. The target system will comprise five identical aluminum cells each 1.25~cm in diameter, 25~cm long, with entrance/exit window thickness of 0.25/0.5~mm and made of aluminum. Four cells will be filled with \nuc{1}{H}, \nuc{2}{H}, \nuc{3}{H}, and \nuc{3}{He} gas while the fifth cell will be empty. We will employ the \nuc{3}{H}, \nuc{3}{He} and empty target cells for the measurement of the \Trit:\He\ cross section ratio. One gas cell will contain 200~psi of tritium, for a density of $\approx$3~mg/cm$^3$, which corresponds to 1~kCi (considerably less than previous experiments). The \nuc{3}{He} cell will be maintained at approximately 350~psi (also $\approx$3~mg/cm$^3$). The target thickness of the \Trit\ cell should be known to about 3\%, and the other cells to \textless~2\%. We also plan to collect data from the other gas cells, as well as a carbon target, to obtain additional cross section ratios and relative normalizations to known cross sections.

The only new piece of equipment will be a collimator plate designed to reduce the number of electrons reaching the spectrometer and to match the event rate across the focal plane. Details regarding the dimensions of the collimator plate and the expected count rates are given in the next section.

\subsection{Extraction of Radius and Corrections}

Extracting the relative charge radius from the cross section ratio includes several corrections. First, to extract the electric form factor $G_E$, the magnetic form factor $G_M$ must be subtracted (Eq.~1). The ratio $\tau$/$\epsilon$ in the reduced cross section is less than 0.3\% for both targets, which implies that the contribution from $G_M$ ($\approx \mu G_E$) to the reduced cross section will be no greater than $\sim$2\%, 
assuming $G_M$/$\mu$ $\approx$ $G_E$. The magnetic radii have been measured to better than 10\% \cite{amroun94}, while calculations quote a much smaller uncertainty \cite{piarulli13}.  Assuming $G_M$ is known to 20\% at the relevant $Q^2$ values, the uncertainty associated with the subtraction of the magnetic contribution is less than 0.4\% for both \Trit\ and \He.

The Coulomb interaction effectively increases the momentum transfer by
\begin{equation}
  Q_{\rm eff} = Q \left( 1 + \frac{3}{2} \frac{Z\alpha \hbar c}{R_{eq}E} \right),
\end{equation}
where $R_{eq}$ is the radius of the nucleus. The relative difference between the effective momentum transfer ($Q_{\rm eff}$) and $Q$ is \textless 0.3\%. In this experiment, it is expected that the uncertainties arising from the contribution of the Coulomb correction will be well-understood and controlled.

In addition to the Coulomb interaction, two photon exchange (TPE) corrections are also considered. Estimates for the correction in \He($e$,$e^\prime$) are $\ll$ 1\% for $Q^2$$\sim$0.05~GeV$^2$ \cite{blunden05,arrington11}. Corrections for scattering from \Trit\ should be comparable and are expected to at least partially cancel in the cross section ratio. It's estimated that the uncertainty from TPE corrections will be $\textless$ 0.3\%.

\section{Rate estimates}

The following section summarizes the rate estimates for the proposed experiment as determined from the latest version of \texttt{MCEEP} \cite{mceep} using form factor parameterizations from \cite{amroun94}. 

\begin{figure}
\setlength{\epsfxsize}{0.40\textwidth}
\leavevmode
\begin{center}
\epsffile{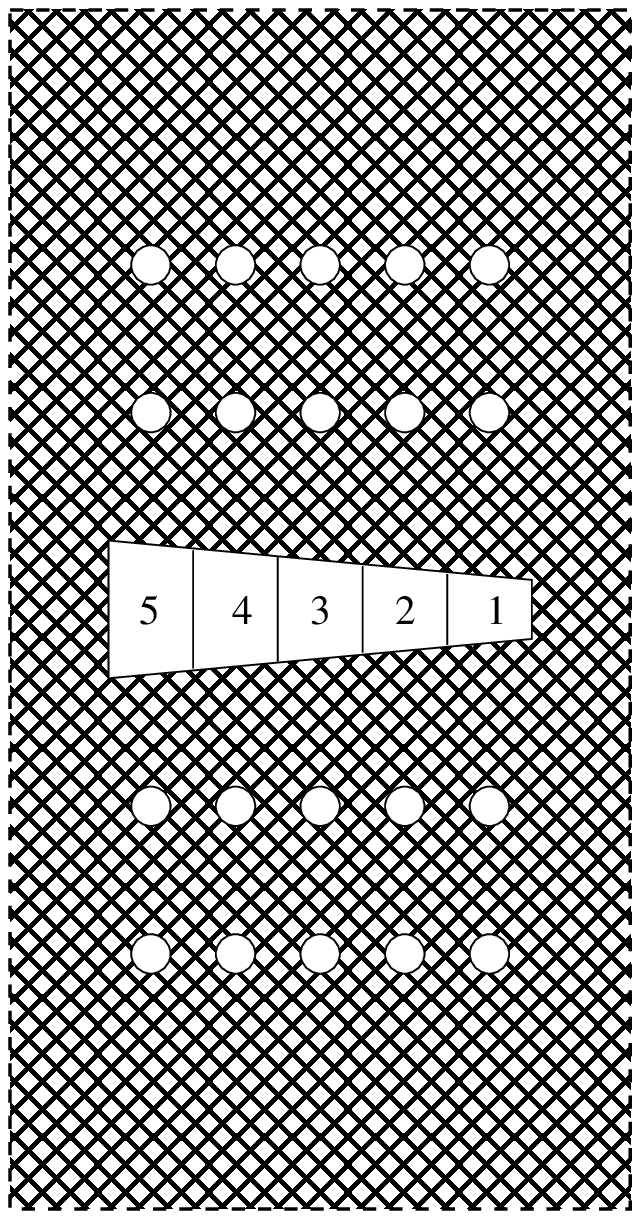}
\end{center}
\caption{The collimator plate design for this experiment. The dashed, outer line indicates the ``standard'' acceptance ($\pm$3.15~cm by $\pm$6.09~cm) for the HRS. The holes are not to scale; they have been enlarged to be visible.}
\label{fig1}
\end{figure}

Using the standard acceptance of the HRS produces a count rate of $\sim$10(40)~kHz from the \Trit(\He) target at 5~$\mu$A of beam. A similar event rate is also anticipated to arise from the Al windows of the target cell. These rates are unsuitable for the HRS in Hall A. Rather than using large pre-scale factors in the DAQ, a special collimator plate will be installed (see Fig.~\ref{fig1}). The plate is designed to match the elastic count rate across the collimator while simultaneously keeping the elastic rate from either target below 3~kHz. Count rates for elastic scattering from both \He\ and \Trit\ are given in Table~\ref{table1}. These are raw rates -- they do not include cuts, radiative corrections, or detector efficiency.

\begin{table}
\begin{center}
\caption{Anticipated elastic scattering count rates for \Trit\ and \He\ at 1.1~GeV. Rates are based on \texttt{MCEEP} simulations with the proposed collimator plate. Note that all $Q^2$ values in the upper half of the table are obtained with the spectrometer positioned at 12.5$^\circ$ and those in the lower half at 15.0$^\circ$. Also, the momentum setting of the HRS does not change. }
\begin{tabular}{ccc|rrrr}
\hline \hline
E$_{\rm beam}$ & $\theta_{\rm HRS}$ & $p_{\rm HRS} $ &  Coll. Bin & $Q^2$  &  \Trit\ Rate  & \He\ Rate \\
 \ [GeV] & [deg] & [GeV/c] &  &  [GeV$^2$]  &  [Hz]  &  [Hz] \\
\hline
 & & & 1  &       0.049 &       180  &  430  \\
 & & & 2  &       0.053 &       220  &  530  \\
1.1 & 12.5 & 1.07 & 3  &       0.057 &       240  &  580  \\
 & & & 4  &       0.061 &       230  &  540  \\
 & & & 5  &       0.065 &       200  &  460  \\
\hline
 & & & 1  &       0.072 &        46  &  100  \\
 & & & 2  &       0.077 &        59  &  120  \\
1.1 & 15.0  & 1.07 & 3  &       0.081 &        69  &  140  \\
 & & & 4  &       0.086 &        67  &  140  \\
 & & & 5  &       0.091 &        57  &  120  \\
\hline \hline
\end{tabular}
  \label{table1}
\end{center}
\end{table}

The event rate from the Al windows of the target cells is expected to be comparable to that from \Trit\ and \He\ ($\sim$2000~Hz at 12.5$^\circ$ and $\sim$200~Hz at 15.0$^\circ$). In addition, contributions from quasi-elastic scattering will increase the rates above those listed here. We anticipate a need to pre-scale the HRS triggers by a factor of 4 with the HRS at 12.5$^\circ$ and a factor of 1 at 15.0$^\circ$.

\begin{figure}
\setlength{\epsfxsize}{0.90\textwidth}
\leavevmode
\begin{center}
\epsffile{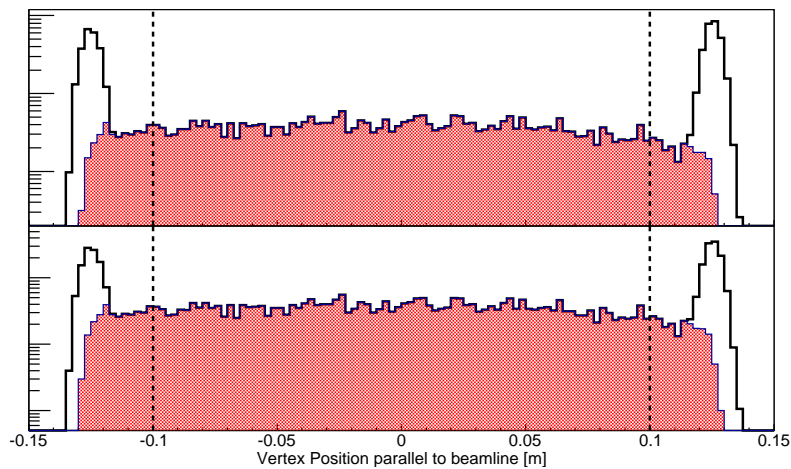}
\end{center}
\caption{The expected number of counts from the target and cell (black) and from the target alone (shaded) as a function of the scattering vertex along the beamline direction. \Trit\ is shown in the top panel and \He\ in the lower. The dashed vertical lines represent software cuts at $\pm$10~cm to remove the contribution from the windows.}
\label{fig2}
\end{figure}

As seen in Fig.~\ref{fig2}, the contributions from the windows can be dramatically reduced by placing a cut on the vertex position at $\pm$10~cm along the beamline. This cut reduces the count rates from the windows to $\sim$0, while the rates for \Trit\ and \He\ shown in Table~\ref{table1} are reduced by $\sim$15\%. Even so, we plan to take data on the empty target to subtract any stray events that might survive the vertex cut during the experimental running.

Additional \texttt{MCEEP} simulations indicate that energy losses and radiative corrections may reduce the elastic count rate by $\sim$20\%. Including this reduction, as well as the reduction in the elastic rate due to the pre-scale factor, a 20\% reduction due to deadtime and 10\% loss due to other effects, the event rate for \Trit\ in each bin would still be $\sim$20~Hz. Using this conservative estimate, one hour of beam on target would produce nearly 10$^5$~counts per bin. Given that the dominant source of background comes from the target-cell windows, and this background can be managed, the statistical uncertainty should be less than the systematic uncertainty arising from the target thickness.

The final experimental uncertainty is due to sources which impact the cross section ratio as well as those which impact the extraction of the $G_E$ ratio. The major source of uncertainty is expected to be the \Trit\ thickness arising from the target pressure. Current estimates place this uncertainty at $\sim$3\%. However, the E12-10-103 experiment will use low-x, deep inelastic scattering data to better constrain the relative target thicknesses. The projected uncertainty in the relative target thickness is 1.5--2\%.

Due to the high count rates, we also propose to collect data from the \D\ and \Hy\ targets as well as a \nuc{12}{C} foil during the beam time. The data from the \nuc{12}{C} and \Hy\ targets will serve as a check of the absolute normalization of the experiment. The \D\ and \Hy\ targets require little overheard or additional beam time and will provide additional comparisons to \Trit\ and \He. All isotopes will be measured relative to \nuc{12}{C} which is well-known and, if necessary, will allow for absolute cross sections to be extracted. Since this unique target system will only be available to Jefferson Lab for a short time, this may be the only opportunity to collect data from all three hydrogen isotopes and \He\ in the same setup.

\subsection{Anticipated Results}

The anticipated uncertainties in the \Trit:\He\ ratio are given in Table~\ref{errtab}. In addition to the experimental uncertainties, the uncertainties from the corrections that are needed to extract the ratio of the charge radii are also tabulated. A total uncertainty of 2\% is expected for the \Trit:\He\ form factor ratio.

\begin{table}
\begin{center}
\caption{Sources of uncertainty in the \Trit:\He\ radius ratio.}
\begin{tabular}{p{4in}r}
\hline \hline
Statistics & 0.4\% \\
Charge & \textless0.5\% \\
Relative target thickness & 1.5--2\% \\
Dead time, tracking, detector efficiency corrections & \textless0.5\% \\
$G_M$ subtraction & 0.4\% \\
Radiative corrections & 0.5\% \\
Coulomb correction, TPE & 0.4\% \\
\hline
{\bf Total} & {\bf 1.8--2.2\%}\\
\hline \hline
\end{tabular}
  \label{errtab}
\end{center}
\end{table}

\begin{figure}
\setlength{\epsfxsize}{0.90\textwidth}
\leavevmode
\begin{center}
\epsffile{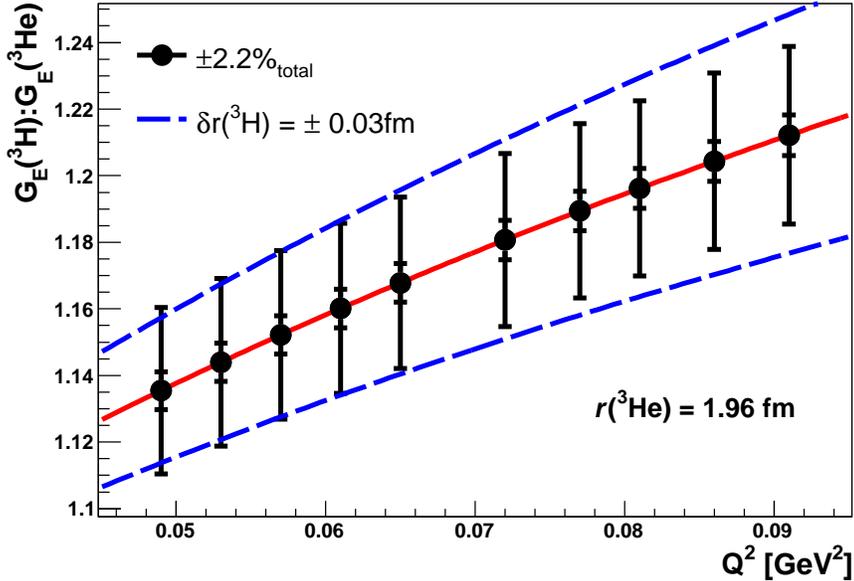}
\end{center}
\caption{Anticipated results of the \Trit:\He\ form factor ratio assuming that $G_E$ has a dipole form. The small error bars represent the statistical uncertainty, and the large are the total uncertainty of 2.2\%. The red curve represents the fit to the data assuming the charge radii reported in Amroun {\it et al.}. The blue curves are the calculated ratio assuming that the \Trit\ charge radius is varied by $\pm$0.03~fm.}
\label{fig3}
\end{figure}

The expected results are shown in Fig.~\ref{fig3}. These data points were generated by assuming that the form factor of both nuclei has a dipole form and using values of $\langle r^2 \rangle$$_{\He}$ = 1.96~fm and $\langle r^2 \rangle$$_{\Trit}$ = 1.76~fm. To investigate the sensitivity to the \Trit\ radius, the \He\ radius was held fixed (as it is well-known through isotopic shifts measurements) and the \Trit\ radius was varied by $\pm$0.03~fm. 

\begin{figure}
\setlength{\epsfxsize}{0.90\textwidth}
\leavevmode
\begin{center}
\epsffile{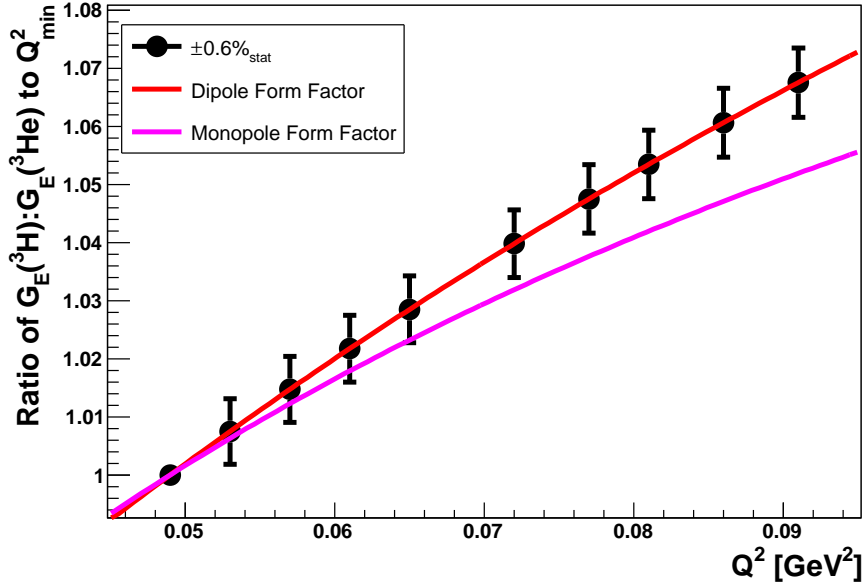}
\end{center}
\caption{The expected ratio of the form factor taken relative to the minimum $Q^2$ point. The uncertanties are statistical only. The systematic effects that dominate the uncertainty in Fig.~\ref{fig3} largely cancel here. The data are generated by assuming that $G_E$ is of a dipole form (red line). The best fit using a monopole form for $G_E$ (magenta) clearly does not match the data trend. It is expected that this type of analysis will constrain the acceptable forms for $G_E$ hence reducing the model uncertainty.}
\label{fig4}
\end{figure}

Additional uncertainty in the extraction of the charge radius can come from the model-dependence of the form factor. To reduce this uncertainty, the high-statistics nature of the data can be exploited to distinguish between various functional forms of $G_E$ using the behavior of the form factor ratio as a function of $Q^2$. As seen in Fig.~\ref{fig4}, the ratios, relative to the minimum $Q^2$ point, can be measured without the large systematic uncertainty arising from the target thickness. The lines shown represent the best fit to the data assuming the $G_E$ has a dipole (red) and monopole (magenta) functional form. In this way, data over the full range of $Q^2$ will reduce the uncertainty due to the modeling of the form factors.

At present, R$_{\He}$-R$_{\Trit}$~=~0.20$\pm$0.10~fm \cite{amroun94}. A 2\% measure of the relative cross section yields an uncertainty in R$_{\He}$-R$_{\Trit}$ of $\pm$0.03~fm. Therefore, we anticipate improving the uncertainty in difference of the radii by more than a factor of three.

\section{Beamtime request}
\label{beamtime}

We request one and a half PAC days of beam time to make the proposed elastic scattering measurement.  As seasoned users of Jefferson Lab, we fully appropriate that a very short experiment can get squeezed by unforeseen events. Never the less, we envision working with the physics division liaison and accelerator program deputy to find a period of time that maximizes our likelihood for a successful short run.  In particular, we would likely try to start setting up our experiment during an accelerator beam study.

In this scenario, the start of the beam study would give us time to access the Hall and get the spectrometer to 12.5$^\circ$ and install our special external sieve while accelerator is doing other work. At the end of the accelerator's beam study, the machine could be scaled to 1.1~GeV (a process that take of order a shift) and we would start our measurements. This assumes that calibration of the beam charge monitor (BCM) and position monitor (BPM) have been completed by the E12-10-103 and/or E12-11-112 experiments (which will run at $\sim$25~$\mu$A), thus reducing our overhead.

The breakdown of our beam time request is given in Table~\ref{table3}. We allow 4 hours for the accelerator to scale down the beam energy to 1.1~GeV. Two hours assigned for BCM calibration at 5~$\mu$A and luminosity studies. Four hours are intended to study the optics and acceptance of the HRS at each kinematic setting using the sieve and the new collimator. Another two hours are allotted to move the spectrometer from 12.5$^\circ$ to 15.0$^\circ$. Furthermore, target changes are assumed to take a total of two hours. This brings the total overhead time to $\sim$18 hours. Eighteen more hours (1.5 hours per target per kinematic setting) are required to obtain the necessary scattering data. In total, we are requesting 1.5 PAC days for these measurements.

\begin{table}
\begin{center}
\caption[LoF entry]{Allotment of proposed beam time.}
\begin{tabular}{p{3.5in}|r}
\hline \hline
Description & Time\\
\hline
Accelerator scaling to 1.1~GeV & 4~hr\\
BCM calibration and luminosity scans & 2~hr\\
Optics and acceptance studies with collimator & 4~hr\\
Production running at 12.5$^\circ$ (1.5 hrs/target) & 9~hr\\
Target changes at 12.5$^\circ$ & 1~hr\\
$^{\rm 1}$Movement of spectrometer from 12.5$^\circ$ to 15.0$^\circ$ & 2~hr\\
Optics and acceptance studies with collimator & 4~hr\\
Production running at 15.0$^\circ$ (1.5 hrs/target) & 9~hr\\
Target changes at 15.0$^\circ$ & 1~hr\\
\hline
{\bf Total Beam Time Request} & {\bf 1.5 PAC Days}\\
\hline \hline
\multicolumn{2}{p{5.0in}}{$^{\rm 1}$ Movement of the spectrometer will require a special access due to the proximity of the HRS to the beam line and the presence of the \Trit\ target.}\\
\end{tabular}
  \label{table3}
\end{center}
\end{table}

\section{Relation to Other Experiments}

The Low Energy Deuteron Experiments (LEDEX) at Jefferson Lab (E05-004, E05-103) collected elastic data from a variety of nuclei -- \D\, \nuc{6}{Li}, \nuc{12}{C}, Ta -- at $Q^2$ \textless 0.1 GeV$^2$. Nuclear radii are being extracted from these data.
As mentioned previously, this experiment will use the same target as the approved E12-10-103, E12-11-112, and E12-14-011 experiments.

\end{document}